%% file: Main.tex
\documentclass[a4paper,11pt]{article}

\usepackage{jinstpub} 

\usepackage{here}
\usepackage{color}
\input supp-mis.mkii

\title{Design and Implementation of Detector Control System for Muon Forward Tracker at ALICE}

\input{Authorlist.tex}

\abstract{
ALICE is the experiment at the CERN LHC devoted to study heavy-ion collisions.
An upgrade program of the ALICE detector is ongoing toward the LHC Run 3 starting in 2022
together with the upgrade of the data acquisition system and the detector control system (DCS)\@.
One of the main projects of the current ALICE upgrade program is the addition of the muon forward tracker (MFT),
a new silicon pixel detector located at forward rapidity.
In this paper, we describe the DCS of the MFT detector
which is entirely controlled via a finite state machine in a hierarchical system.
}


\begin{document}
\maketitle
\flushbottom

\input{Introduction.tex}
\input{Overview.tex}
\input{Operation.tex}
\input{Implementation.tex}
\input{Summary.tex}


\input{Acknowledgments.tex}


\input{Bibliography.tex}
\end{document}

%% file: Authorlist.tex
\author[a]{K.~Yamakawa,} 
\author[b]{A.~Augustinus,}
\author[c]{G.~Batigne,}
\author[b]{P.~Chochula,}
\author[a]{M.~Oya,}
\author[d]{S.~Panebianco,}
\author[b,e]{O.~Pinazza,}
\author[a,1]{K.~Shigaki,\note{Corresponding author.}}
\author[f]{R.~Tieulent,}
\author[a]{and Y.~Yamaguchi}

\affiliation[a]{Hiroshima University, Hiroshima, Japan}
\affiliation[b]{European Organization for Nuclear Research (CERN), Geneva, Switzerland}
\affiliation[c]{SUBATECH, IMT Atlantique, Universit\'{e} de Nantes, CNRS-IN2P3, Nantes, France}
\affiliation[d]{Universit\'{e} Paris-Saclay Centre d'Etudes de Saclay (CEA), IRFU, D\'{e}partment de Physique Nucl\'{e}aire (DPhN), Saclay, France}
\affiliation[e]{INFN, Sezione di Bologna, Bologna, Italy}
\affiliation[f]{Universit\'{e} de Lyon, Universit\'{e} Lyon 1, CNRS/IN2P3, IPN-Lyon, Villeurbanne, Lyon, France}

\emailAdd{shigaki@hiroshima-u.ac.jp}

%% file: Introduction.tex
\section{Introduction}
\label{sec:Intro}
A Large Ion Collider Experiment (ALICE)~\cite{ALICE} is the experiment
which focuses on the heavy-ion program at the CERN 
Large Hadron Collider (LHC)~\cite{LHC}\@.
Understanding of the properties of quark-gluon plasma (QGP),
which is an exotic state of hadronic matter described by quantum chromodynamics,
is the primary aim of ALICE\@.

\subsection{ALICE Upgrade in LHC Run 3}
\label{subsec:Upgrade}
ALICE has performed successfully
during LHC Runs 1 (2009--2013) and 2 (2015--2018)
harvesting a multitude of results.
To take advantage of the increased luminosity of the LHC 
during Runs 3 (2022--2024) and 4 (2027--2030)
for high precision measurements of experimental observables
and to extend its scientific goals,
the ALICE collaboration defined a complete upgrade strategy~\cite{UPGRADE}.
The aim of the ALICE upgrade is
to have the capability of recording all Pb-Pb interactions
in a continuous data taking mode
and to enhance the track reconstruction performance~\cite{UPGRADE, READOUT, ITS, MFTTDR, TPC, O2}.
The implementation of this upgrade program includes, 
in particular the readout and trigger for the new front-end electronics~\cite{READOUT},  
a new integrated online-offline computing system (O$^{2}$)~\cite{O2}, 
and the addition of the muon forward tracker (MFT)~\cite{MFTLOI, MFTTDR},
a silicon pixel tracker at forward rapidity.

\subsection{Muon Forward Tracker}
\label{subsection:MFT}
Heavy quarks, charm ($c$) and bottom ($b$) quarks, are known
as good probes to investigate the characteristics of the QGP\@.
The muon spectrometer of ALICE~\cite{MUON, MUON2} has performed successful measurements
of the $J/\psi$ production rate in the forward rapidity region
during Runs 1 and 2 in various collision systems from
pp, p-Pb to Pb-Pb~\cite{LHC}.
The separation of $J/\psi$ from B hadron decay from prompt $J/\psi$ has been, however,
impossible due to the presence of a hadron absorber of 60~radiation lengths
placed between the interaction point and the muon tracking system.
It induces large multiple scatterings and limited spatial resolution around the interaction point,
preventing the determination of the origin of muons.
In order to overcome this limitation,
a new silicon pixel detector, the MFT, is installed
between the interaction point and the hadron absorber.
It covers the forward pseudo-rapidity range of -3.6 < $\eta$ < -2.5
to match most of the muon spectrometer acceptance.
The pointing accuracy of the muon production point is consequently improved
by matching the tracks measured by the MFT and by the muon spectrometer.
The CMOS monolithic active pixel sensor (CMOS-MAPS) technology has been chosen for the sensors.
The adopted sensor is developed for both the new ALICE inner tracking system (ITS) and the MFT,
and is called the ALICE pixel detector (ALPIDE)~\cite{ALPIDE,ALPIDE2}.
The dimension of ALPIDE is 1.5 $\times$ 3~cm$^2$ with the pixel pitch of 27 $\times$ 27~$\mu$m$^2$.
It has a spatial resolution of about 5~$\mu$m and the charge integration time of 30~$\mu$s.
A total of 936 ALPIDE chips are used for the MFT covering about 0.4~m$^2$.

\begin{figure}[htbp]
\centering 
\includegraphics[width=6cm]{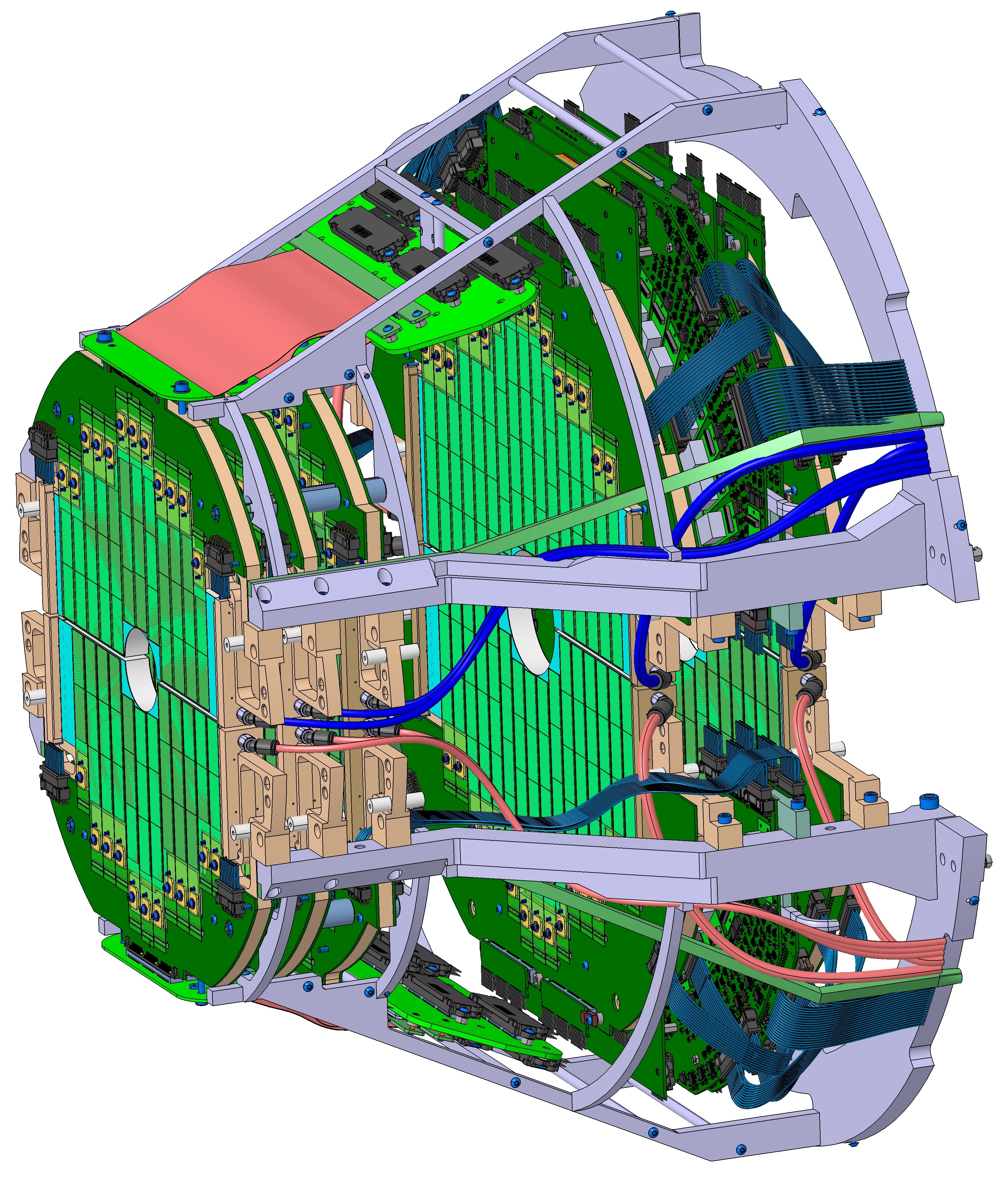}
\caption{\label{fig:MFT} 3-dimensional view of the MFT detector.}
\end{figure}

\begin{figure}[htbp]
\centering 
\includegraphics[width=9cm]{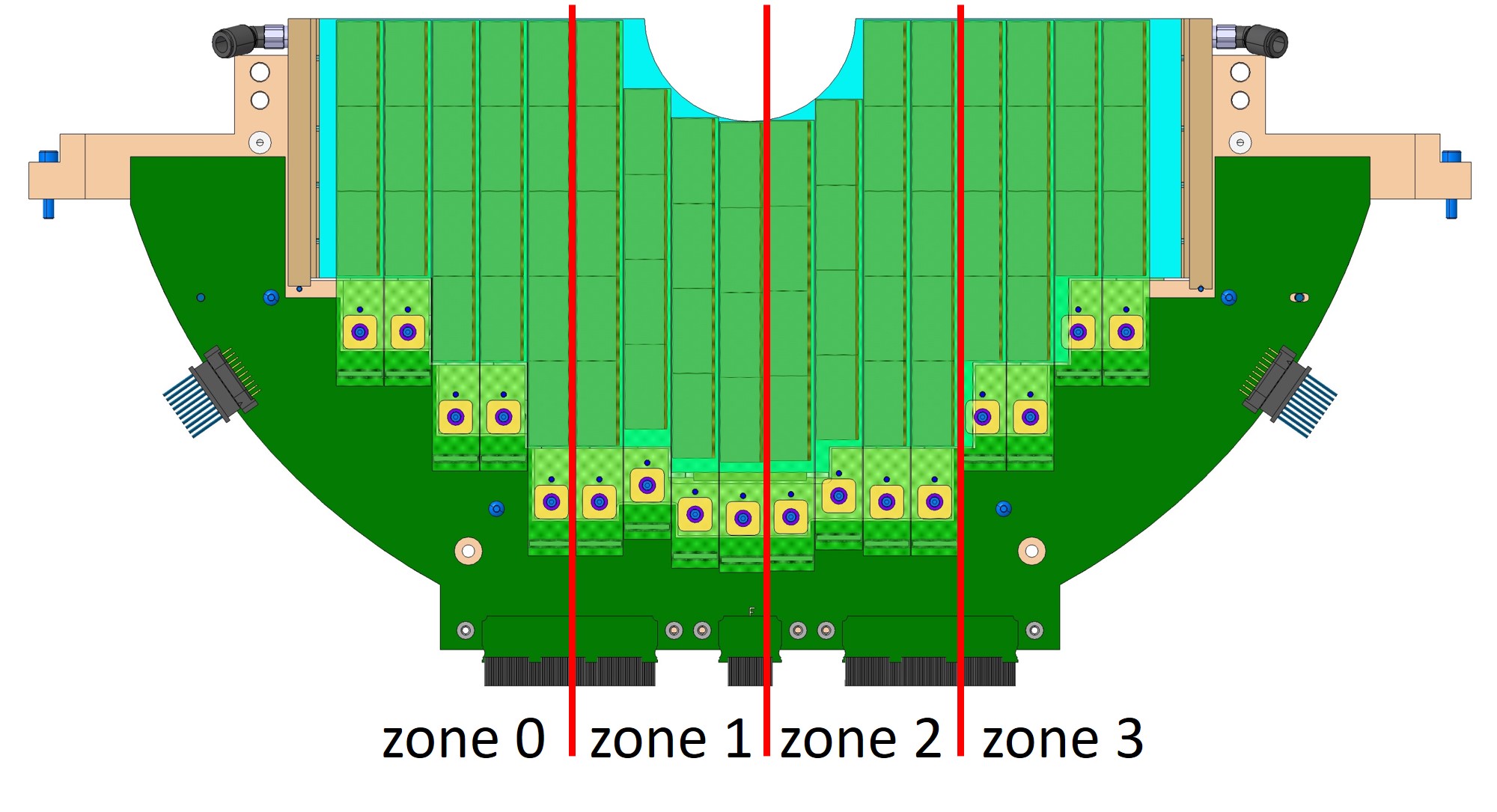}
\caption{\label{fig:Disk} Definition of zones of half disk 4.} 
\end{figure}

Figure~\ref{fig:MFT} shows the 3-dimensional view of the full MFT\@.
The MFT is separated into two half cones, called the top and the bottom MFT's, respectively.
A half cone is composed of five half disks, named from 0 to 4 (\emph{e.g.} half disk 0),
each with two detection half planes.
Each detection half plane is split into four zones,
each of which is powered in common and read out in order to reduce the number of connection lines.
Figure~\ref{fig:Disk} shows the definition of zones of half disk 4 as an example.
A zone corresponds to a set of three to five sensor ladders
connected to a single readout unit board (RU)\@.
Each ladder, housing between two and five ALPIDE chips,
is a flexible printed circuit board (PCB) on which the sensors are wire-connected.
The ladders are glued on the support planes
and connected to another PCB enabling the power and data connection.
A total of 280 ladders composes the full MFT\@.

\subsection{Online-Offline Computing System}
\label{subsec:O2}
Continuous readout of raw data without any trigger and subsequent simultaneous data processing
are a challenge of the ALICE upgrade program.
A typical data volume produced by the ALICE sub-detectors will be 3.4~TB/s
in Pb-Pb collisions at $\sqrt{s_{\bf NN}} =$ 5.5~TeV at a collision rate of 50~kHz.
ALICE hence develops a new computing system, named the O$^{2}$,
in which online and offline systems are merged in a single operating system~\cite{O2}.
The data links between the acquisition system and the front end electronics (FEEs)
use the gigabit transceiver (GBT)~\cite{GBT} technology developed at CERN\@.

\begin{figure}[htbp]
\centering
\includegraphics[width=14cm]{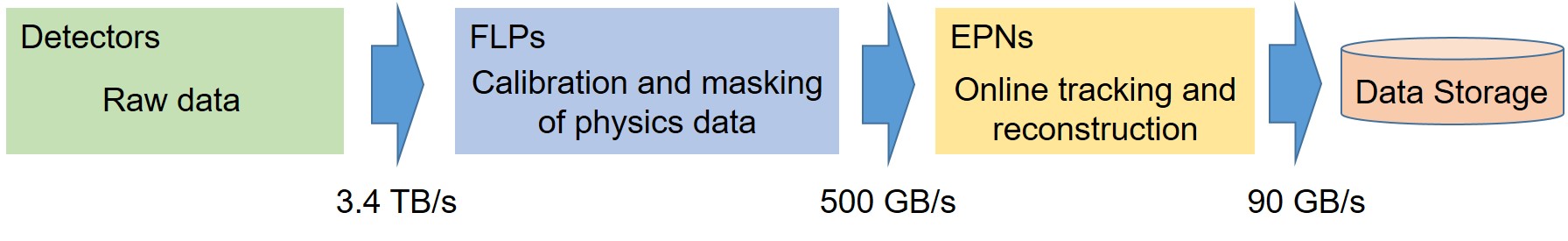}
\caption{\label{fig:O2} Diagram of the O$^2$ data stream.
}
\end{figure}

In the O$^{2}$ farm,
the common readout unit (CRU) on the first level processor (FLP) splits raw data from the sub-detectors
into physics data and slow control data.
Figure~\ref{fig:O2} shows the diagram of the O$^2$ raw data stream.
The slow control data for each event are one of the ingredients for the data reduction on the FLP\@.
Collected physics data on the FLP are transferred to the event processor nodes (EPNs)
with a data rate of 500~GB/s.
The EPN processes the raw data online while performing reconstruction tasks such as clustering and tracking.
The output of these tasks is transmitted to the data storage at a rate of 90~GB/s.

\subsection{Detector Control System}
\label{subsec:ALICEDCS}
\begin{figure}[htbp]
\centering
\includegraphics[width=14cm]{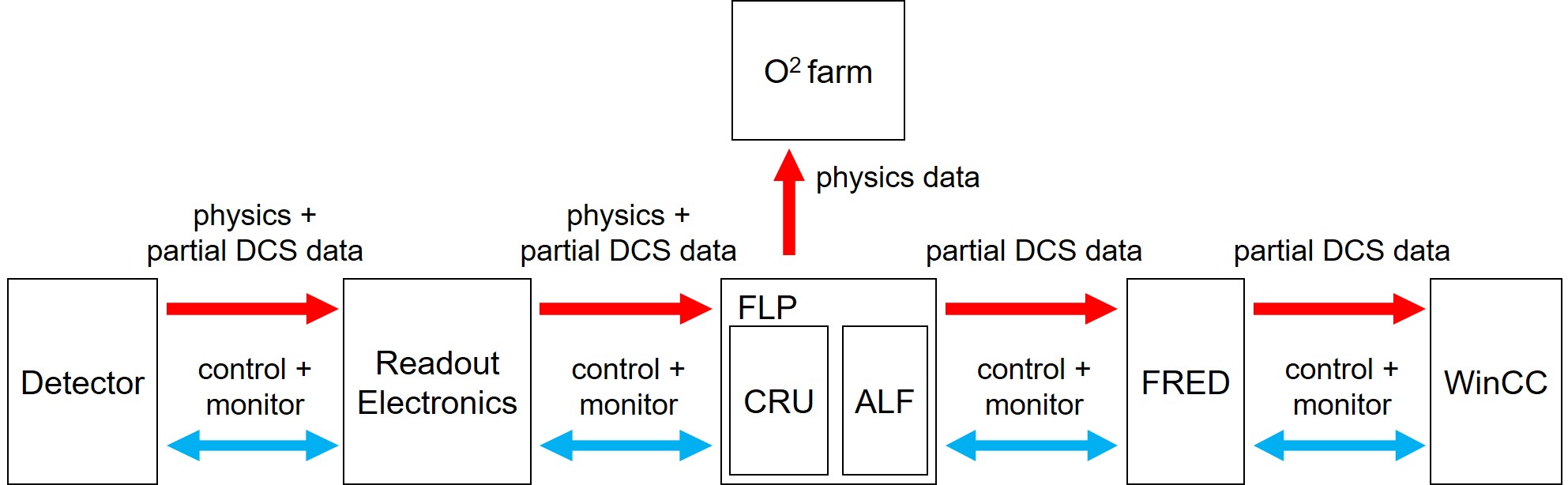}
\caption{\label{fig:DCS} Schematics of the DCS data stream.
Control commands are transmitted from WinCC OA to the MFT through the FRED and the FLP\@.
The monitoring data are collected via the same links in the other way,
as shown with the blue arrows.
A part of DCS data that are temperatures of ALPIDEs share the same packets
with the physics data from the MFT to the CRU. 
The physics data are sent to the O$^2$ farm after splitting from the DCS data at the CRU, 
while the DCS data come to WinCC OA via Ethernet,
as shown with the red arrows.
}
\end{figure}

The detector control system (DCS) of ALICE is upgraded to follow the O$^{2}$ strategy.
It is based on the component framework, guidelines, and configurations
of a framework named the joint control project (JCOP)~\cite{JCOP},
which is developed at CERN and provides software tools for DCS development
on WinCC Open Architecture (WinCC OA)~\cite{WINCCOA}.
The control of FEE employs the GBT slow control adapter (GBT-SCA)~\cite{GBTSCA},
which is an application-specific integrated circuit (ASIC),
designed for slow control in the framework of the GBT\@.
A specific software framework is developed
in order to interface the detector's FEE from the graphical user interface (GUI)
based on WinCC OA\@.
The framework includes two major elements:
the ALICE low-level front-end (ALF) running on the FLP and the front-end device (FRED)~\cite{ALFRED}
as shown in Fig.~\ref{fig:DCS}\@.
The ALF provides low-level access to the CRU links,
while the FRED translates high-level instructions from WinCC OA
into low-level commands consisting of sequences of hexadecimal words
to operate the GBT-SCA and access the FEE\@.
Slow control data, including detector and environmental conditions,
are carried in the same packet with physics data in the raw data stream from the FEE\@.
The CRU splits the raw data into physics and slow control data.
The slow control data come up to WinCC OA via the ALF and the FRED,
while the physics data are transferred to the O$^2$ farm as described above.
The communication protocol between the WinCC OA, the FRED, and the ALF is
based on the CERN distributed information management system (DIM)~\cite{DIM}.

%% file: Overview.tex
\section{Detector Control System for the Muon Forward Tracker}
\label{sec:MFTDCS}

\subsection{Hardware Structure}
\label{subsec:HwStructure}
\begin{figure}[htbp]
\centering
\includegraphics[width=11cm]{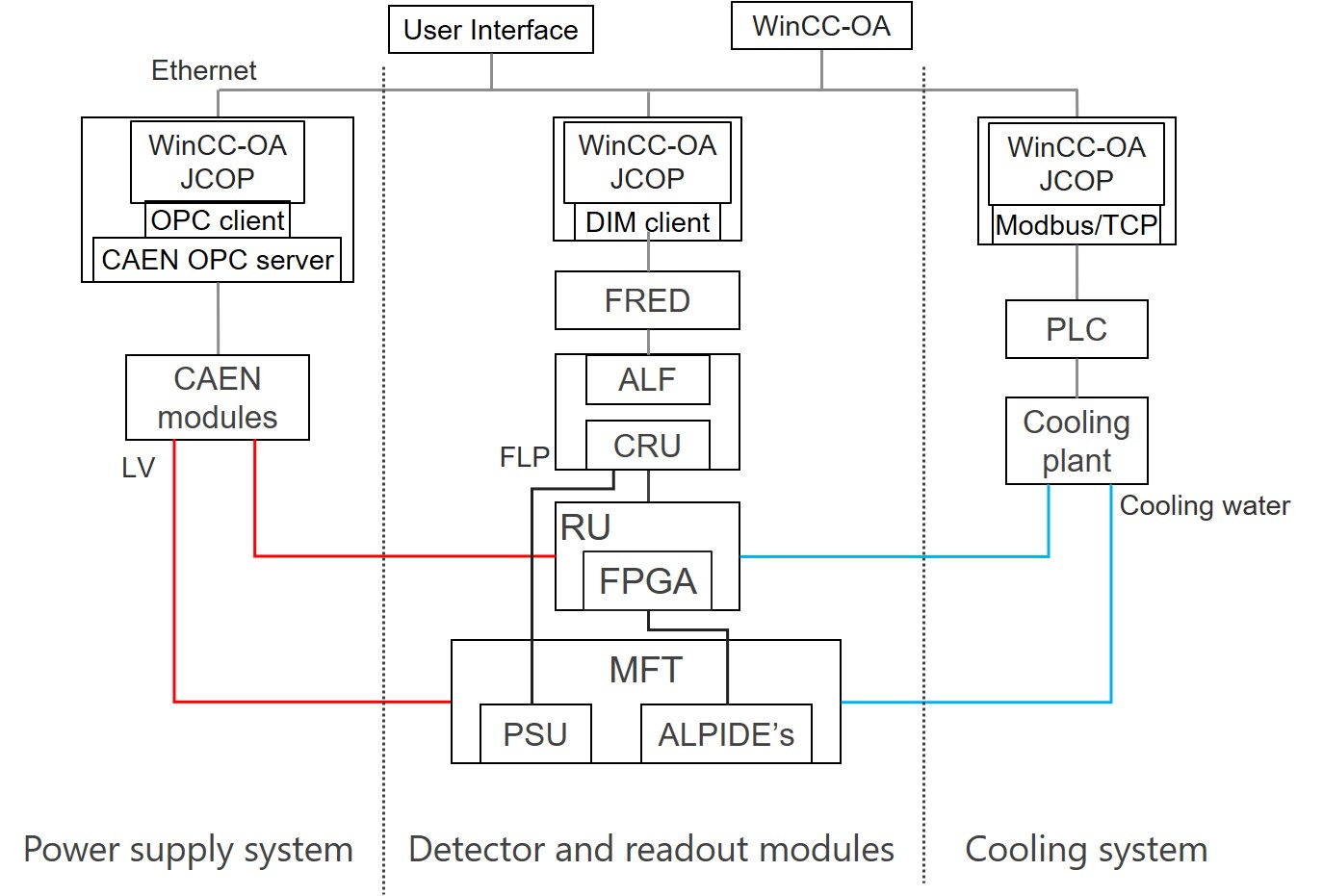}
\caption{\label{fig:URD} Hardware structure of the MFT DCS\@.}
\end{figure}

The MFT DCS controls and monitors three subsystems as shown in Fig.~\ref{fig:URD}:
the low voltage power supplies, the detector and readout modules, and the cooling system.

\subsubsection{Power Supply System}
\label{subsubsec:PS}
The ALPIDE chip is powered by two voltage lines at 1.8~V, for the analog and digital parts.
In addition, a reverse bias voltage up to -3~V can be applied to the ALPIDE sensor
to increase its efficiency and cope with the performance degradation induced by the radiation dose.
A dedicated board, known as the power supply unit (PSU),
is installed inside the detector, between half disks 3 and 4,
to provide local voltage generation
in order to avoid large voltage drops in the power cables from the power supplies to the detector,
located about 40~m apart.
The PSU board houses DC-DC converters providing the 1.8~V outputs to the detector
as well as the GBT-SCA chips to control and monitor the PSU via the CRU\@.

\begin{figure}[htbp]
\centering
\includegraphics[width=14cm]{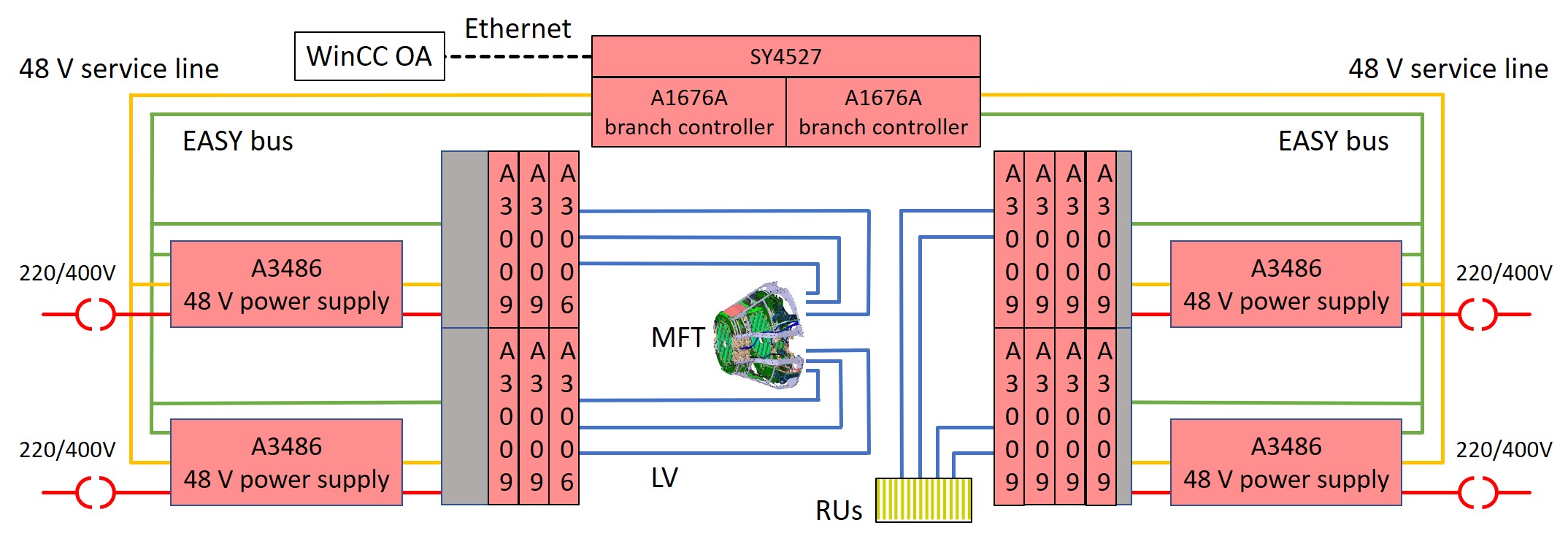}
\caption{\label{fig:PS} Structure of the power supply system.}
\end{figure}

Power supply modules manufactured by CAEN~\cite{CAEN} are used
to supply the low voltage (LV)\@.
Figure~\ref{fig:PS} shows the structure of the LV system of the MFT\@.
WinCC OA connects with an SY4527 mainframe using open platform communications (OPC) via Ethernet.
Two branch controllers A1676A in the SY4527 communicate with two power supply systems,
one to power the PSUs and the other to power the FEE cards, named the readout units (RUs)\@.
These systems are based on the CAEN embedded assembly system (EASY)
which is tolerant to radiation and magnetic field.
Twelve A3009 power supply boards and two A3006 boards are installed in four EASY3000 crates,
which are powered by four A3486 modules to convert 3-phase AC to 48~V DC\@. 

\subsubsection{Detector and Readout Modules}
\label{subsubsec:Readout}
The RU, a FPGA based system, is employed as the FEE card of the MFT
to read raw data from and send configuration to the ALPIDE chips.
One RU board reads out raw data from a zone of a detection half plane.
A total of 80 RUs composes the MFT FEE system.
The FPGA on the RU is used for configuration and operation of the ALPIDE chips.
The RU and the CRU communicate through GBT links.


\subsubsection{Cooling System}
\label{subsubsec:Cooling}
\begin{figure}[htbp]
\centering
\includegraphics[width=14cm]{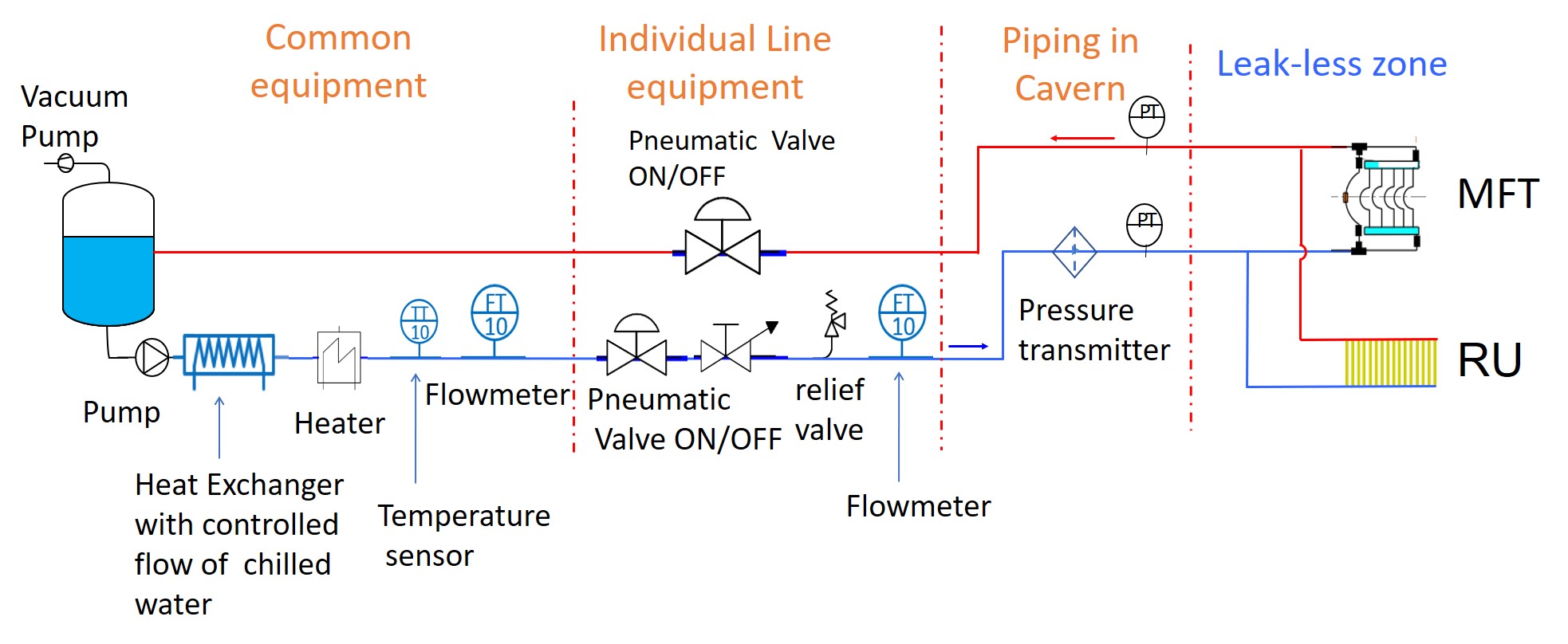}
\caption{\label{fig:Cooling} Structure of the water cooling system.}
\end{figure}

A leakless water system is used to ensure proper cooling of the detector and the RUs.
The structure of the system is shown in Fig.~\ref{fig:Cooling}.
A nominal pressure value of the cooling water is set at 0.3~bar,
below the atmospheric pressure to prevent a water leak.
The temperature ranges of inlet water to the detector and to the RUs are
18--20~\textdegree{}C and 18--23~\textdegree{}C, respectively.

An air ventilation system provides a dry and cool airflow
which guarantees the temperature uniformity and humidity control inside the detector volume.
The nominal values of temperature and humidity are 20~\textdegree{}C and 35$\%$, respectively.

\subsection{Logical Structure}
\label{subsec:LogicalStructure}
\begin{figure}[htbp]
\centering
\includegraphics[width=10cm]{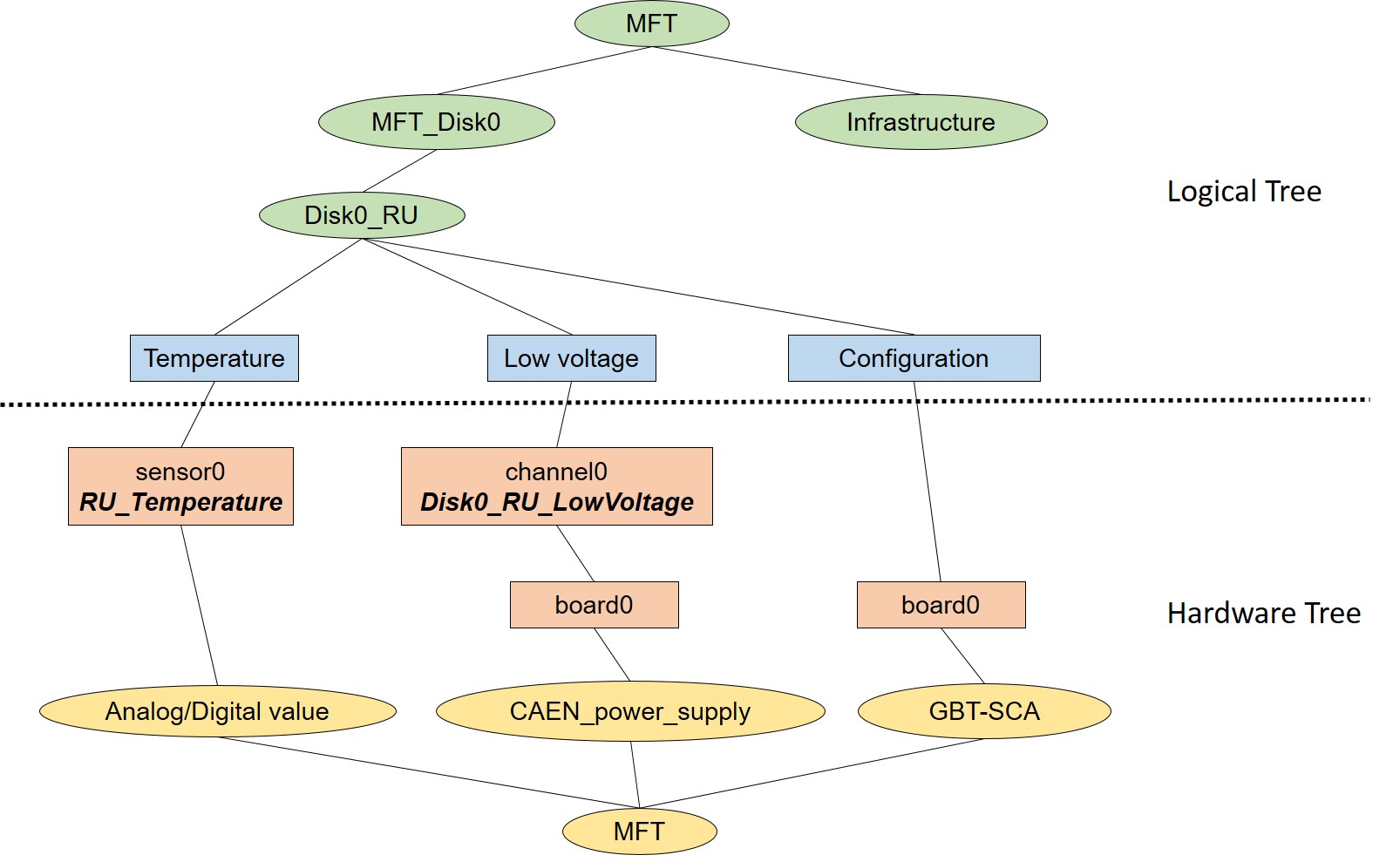}
\caption{\label{fig:Logical}
Relation between the logical tree and the hardware tree of the MFT DCS\@.
Names written in italic are aliases of the hardware device for the logical tree.
}
\end{figure}

A logical tree structure, 
which describes all devices in operation and monitoring,
is designed based on the hardware structure of the MFT DCS,
and configured using the JCOP device models.
On the one hand, all hardware devices are referred as elements of the tree,
and control commands are published from the elements to the devices.
On the other hand, the hardware tree on WinCC OA describes how all hardware devices are wired up.
Figure~\ref{fig:Logical} shows the relation between the logical tree and the hardware tree.

The logical representation of the detector is built by using aliases
on the JCOP device instances.
An alias name is assigned to each device used in operation,
and reassigned to point a different hardware channel,
for example when the originally corresponding low voltage power supply channel has a problem.

%% file: Operation.tex
\section{Operation and Interlock}
\label{sec:Opearation}

\subsection{Finite State Machine}
\label{subsec:FSM}
\begin{figure}[htbp]
\centering 
\includegraphics[width=15cm]{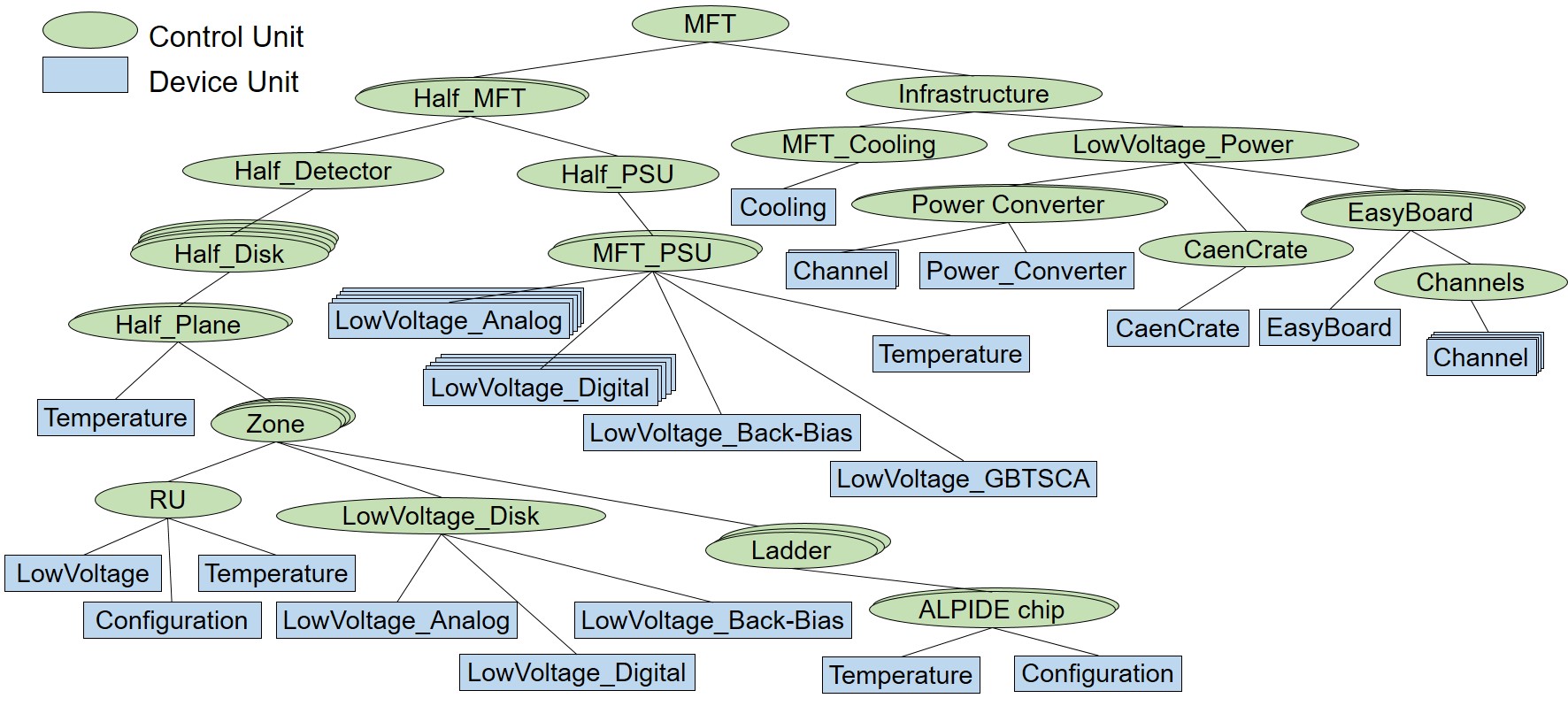}
\caption{\label{fig:FSM_Tree} FSM structure of the MFT DCS\@.
It is designed based on the tree-like logical representation.}
\end{figure}

The finite state machine (FSM) of the MFT detector
is a hierarchical control application,
based on its tree-like logical representation and state diagrams
implemented in the JCOP framework on WinCC OA\@.
Figure~\ref{fig:FSM_Tree} shows the FSM structure of the MFT DCS\@.
A control unit is a conceptual part of the detector system
and a device unit corresponds to a real device controlled by the FSM\@.
The FSM allows modeling of behaviors of the elements with state diagrams.

\begin{figure}[htbp]
\centering
\includegraphics[keepaspectratio, width=13cm]{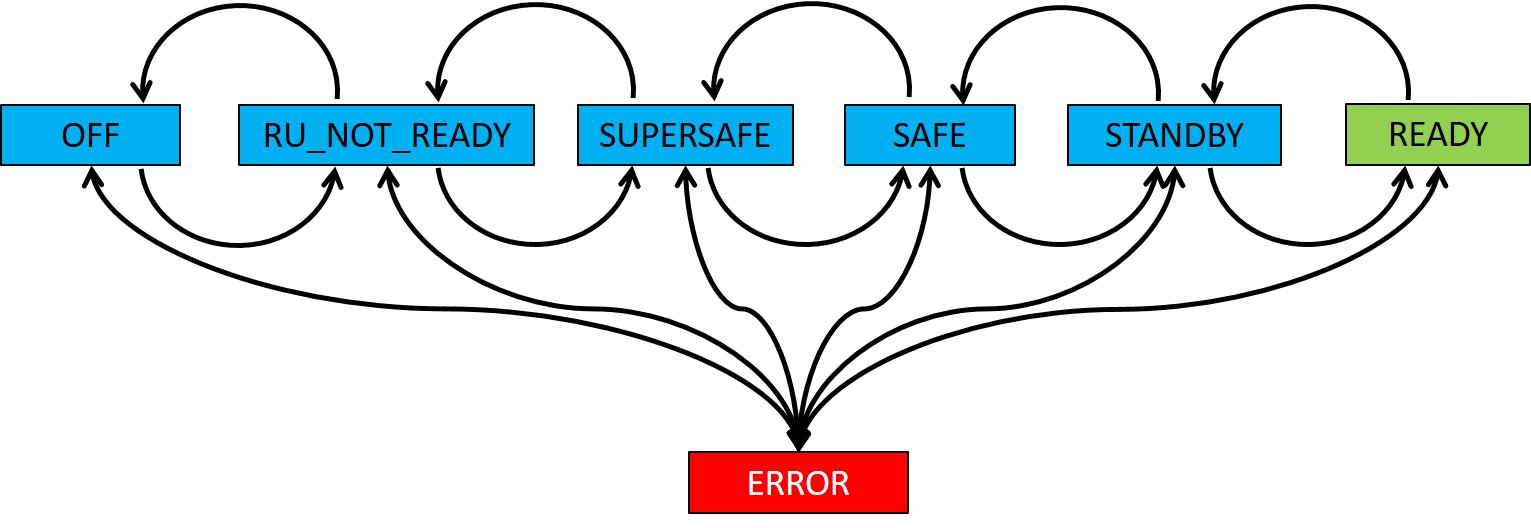}
\caption{\label{fig:FSM_State}
State diagram of the MFT top node.
The MFT detector is available for physics data taking only in the green READY state.
The states with blue boxes mean the status is okay, but not ready for data taking.}
\end{figure}

\begin{figure}[htbp]
\centering
\includegraphics[keepaspectratio, width=13cm]{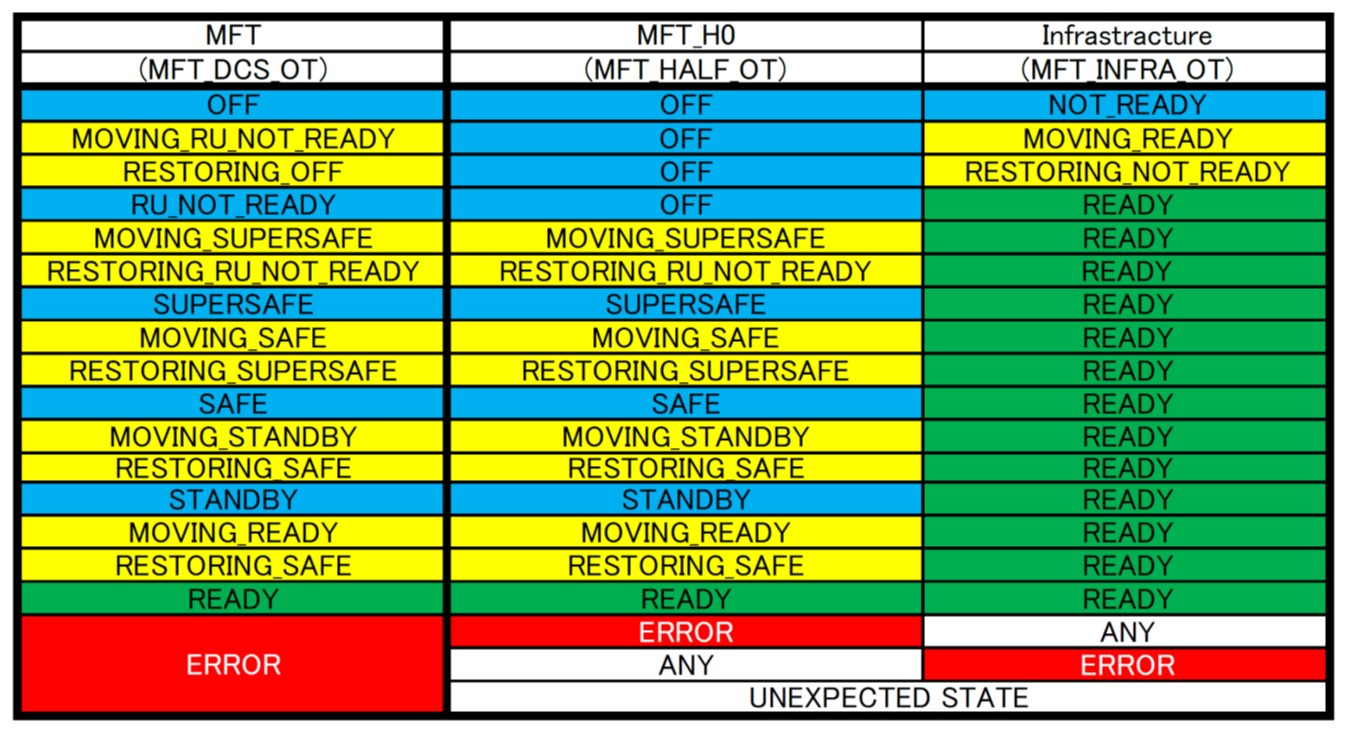}
\caption{\label{fig:FSM_Table} Synchronization table of the MFT top node.}
\end{figure}

The FSM as a software artifact
propagates commands across the devices and sums up the overall state of the detector.
State diagrams are defined for all the FSM nodes.
Figure~\ref{fig:FSM_State} shows the diagram for the top node, named MFT in Fig.~\ref{fig:FSM_Tree},
as an example.
A series of actions permits to switch on/off and configure the different parts of the detector system
according to a given sequence.
Figure~\ref{fig:FSM_Table} is the synchronization table
which shows how the top node state is defined based on the states of the two daughter nodes,
corresponding to the related subsystems.
The only state which allows for physics data acquisition is the READY state
where all subsystems are on and configured.
Other states include
intermediate states used either to bring up/down the detector between the OFF and READY states,
and secured states for special conditions
of the LHC beam (e.g.\ magnets ramping, beam tuning, beam injection, and beam dumping)
or of the ALICE experiment (e.g.\ changing magnet conditions).
The state of the top node moves from any state to the ERROR state whenever a problem arises.

\paragraph{Operation}
FSM commands are transferred down the hierarchy and converted into slow control commands,
which are in turn transferred to the ALPIDE chips, the PSU, and the RU through DIM\@.
The commands are sent from the FSM also to the CAEN power supply modules via the EASY bus. 
The upper nodes of the FSM send operational commands to their daughter nodes,
then the lowest control units transmit the operational commands to the end devices,
which the device units correspond to\@.
In the other direction,
condition data from the ALPIDE chips and the RU pass through the DCS data line to the FSM\@.

\paragraph{Software Interlock}
The FSM implements an automatic software interlock mechanism.
The FSM monitors the temperatures of the GBT-SCA, the half planes,
the RU, the FPGA on the RU, the mezzanine boards of the PSU, and the ALPIDE chips.
When one of the monitored temperatures exceeds a given threshold,
the FSM turns off the corresponding channel of the CAEN LV module.
The flow and humidity of the cooling air and the temperature of the half disks are also monitored by the FSM\@. 
All the LV power supply channels for the entire MFT detector are turned off
if the FSM detects an abnormal condition of the cooling air
or an excessive temperature of a half disk.

The FSM also takes care of communication control to the CAEN mainframe.
It refreshes the OPC server for the CAEN modules
when the communication between the CAEN mainframe and WinCC OA is lost.
Communication loss between WinCC OA and the FRED and/or the FLP is another case to trigger the software interlock.

\subsection{Detector Safety System}
\label{subsec:DSS}
\begin{figure}[htbp]
\centering 
\includegraphics[width=13cm]{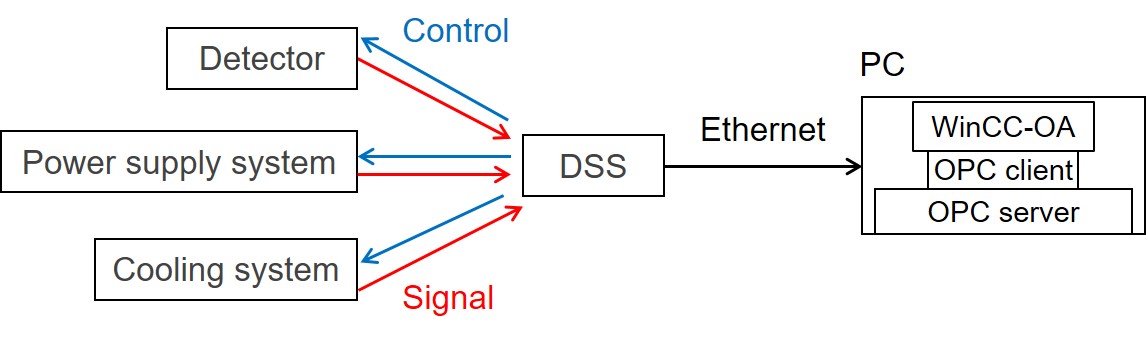}
\caption{\label{fig:DSS}
MFT Detector Safety System (DSS) layout.}
\end{figure}

Detector safety system (DSS) is the hardwired interlock system
based on programmable logic controllers (PLCs),
used commonly for all sub-detectors in ALICE\@.
Actions on the DSS are implemented detector by detector.
Figure~\ref{fig:DSS} shows the general layout of the MFT DSS\@.
It turns off all channels of the CAEN power supply modules for the detector
if a crucial problem occurs.

In this multi-layered scheme combining the DSS and the preventive software interlock by the DCS FSM,
minor issues, {\it e.g.}\ an excessive temperature in the MFT
or a communication loss with the main power supply,
are normally handled by raising an alarm and shown in the FSM\@.
The DSS represents an ultimate safety system in these cases,
being operational even if the communication
between the WinCC OA and the CAEN main frames is lost.
Any issue in the cooling system should be handled directly by the DSS\@.
The sensors on the cooling system hence give a hardwired trigger to the DSS
in case the cooling does not work correctly.

%% file: Implementation.tex
\section{Implementation and Tests with Detector Hardware}

\subsection{Ladder Smoke Test}
\label{subsec:SmokeTest}
A quality assurance system of the ladders has been set up
during the ladder production phase of the MFT project.
This system also serves as a basis to develop
the first elements of the global MFT DCS system on WinCC OA\@.
The first step of the MFT ladder qualification is to power it.
This test is named a smoke test.
The voltages provided to the analog and digital parts of the chips are ramped up
from 0~V to the nominal value of 1.8~V by steps of 0.1~V\@.
The consumption of current is recorded
and any abnormal power consumption triggers a voltage shutdown.
The smoke test detects some defects of the ladders,
in both conditions with and without the back-bias voltage.

\begin{figure}[htbp]
\centering 
\includegraphics[width=15cm]{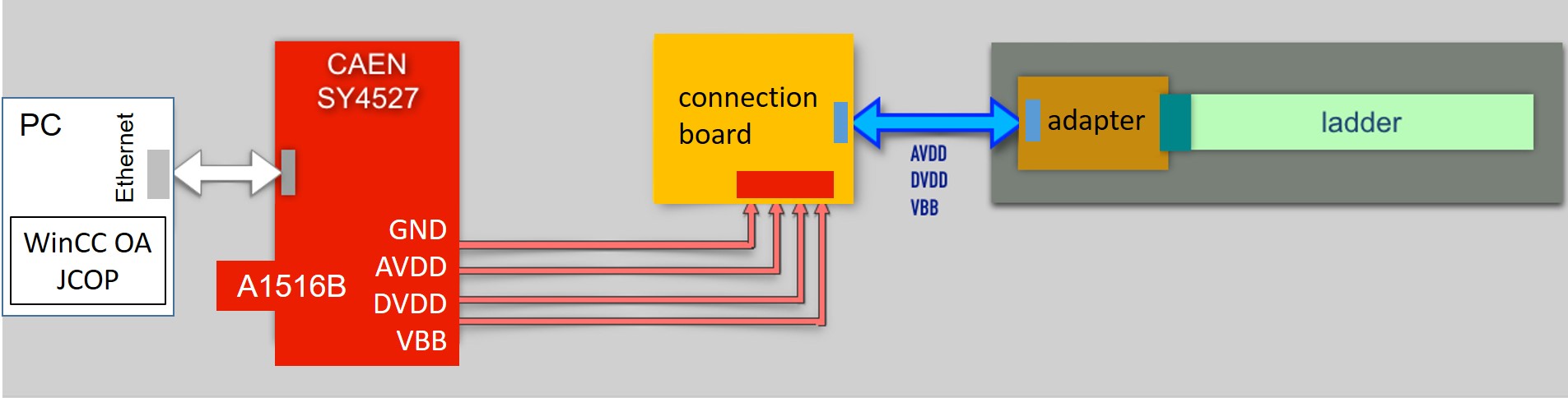}
\caption{\label{fig:Smoke_test} Smoke test bench setup.}
\end{figure}

\paragraph{Setup}
Figure~\ref{fig:Smoke_test} shows the setup of the smoke test bench.
It consists of a WinCC OA instance including the JCOP framework,
the CAEN power supply system,
and intermediate boards to adapt the cable connection to the ladder.
An A1516B low voltage power supply board is inside an SY4527 crate. 
It supplies voltages in the required ranges of 0.0--1.8~V for analog and digital lines
and of -3.0--0.0~V for the back-bias.
An application running under WinCC OA controls the power supply system and records the test results. 
A GUI has been designed and implemented on WinCC OA\@.
Operators can set the demanded values of output voltages,
the numbers of steps between 0.0~V and the demanded values,
and the ramping up speed in each of the steps.
The test results are recorded in comma separated values (CSV) files
and screen shots of the GUI in portable network graphical format (PNG) files as the logs.
A safety system is implemented to protect the ladder. 
The test is stopped when the current exceeds a given threshold
or if the GUI is closed.

\begin{figure}[htbp]
\centering 
\includegraphics[width=10cm]{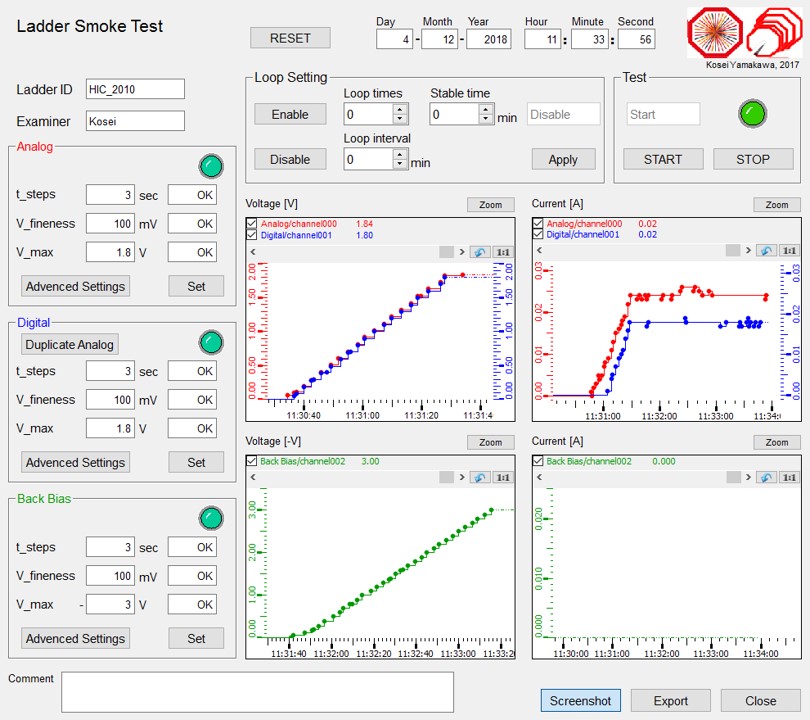}
\caption{\label{fig:SmokeTest_Result} Result of smoke test with back-bias voltage.}
\end{figure}

\paragraph{Achievements}
The smoke test needs to be performed for about 500 ladders including spares.
Figure~\ref{fig:SmokeTest_Result} shows the test results of one of the ladders,
numbered 2010.
A test is performed first without the back-bias voltage.
The current consumption of the analog and digital lines reaches
a constant value at about 20~mA each
which is below the maximum allowed value of 40~mA\@.
The negative back-bias voltage is then slowly increased from 0.0 to -3.0~V
keeping analog and digital voltages at the nominal values.
The power consumption of the analog and digital lines should stay constant
and the back-bias current should not exceed 20~mA\@.
In this example, the smoke test is successful and the ladder is made available
for the next steps of the qualification procedure
before being used for the MFT detector assembly.

\subsection{MFT Surface Commissioning}

The MFT detector is commissioned at CERN,
following the production and assembly of its components.
The DCS is integrated to the detector system in this commissioning phase
in order to test
all its functionalities.
This stage is called the surface commissioning,
{\it i.e.} before the MFT installation in the cavern where the ALICE detector is located.
Extensive readout tests are conducted during this stage.
The MFT DCS is crucial in this phase
to ensure the safe and easy operation.

The MFT DCS consists of three separate subsystems as described in Sec.~\ref{sec:MFTDCS}.
The entire MFT detector system, except for the cooling part, was assembled during the commissioning.
A WinCC OA project is dedicated for each of the subsystems
and they are integrated by the main instance known as the MFT DCS through Ethernet.
Relevant data items in terms of the FSM nodes are
put together in a single DCS GUI panel.
The panels are refined to allow more user-friendly operations
through the MFT surface commissioning stage.

\begin{figure}[htb]
  \centering 
  \includegraphics[width=0.6\linewidth]{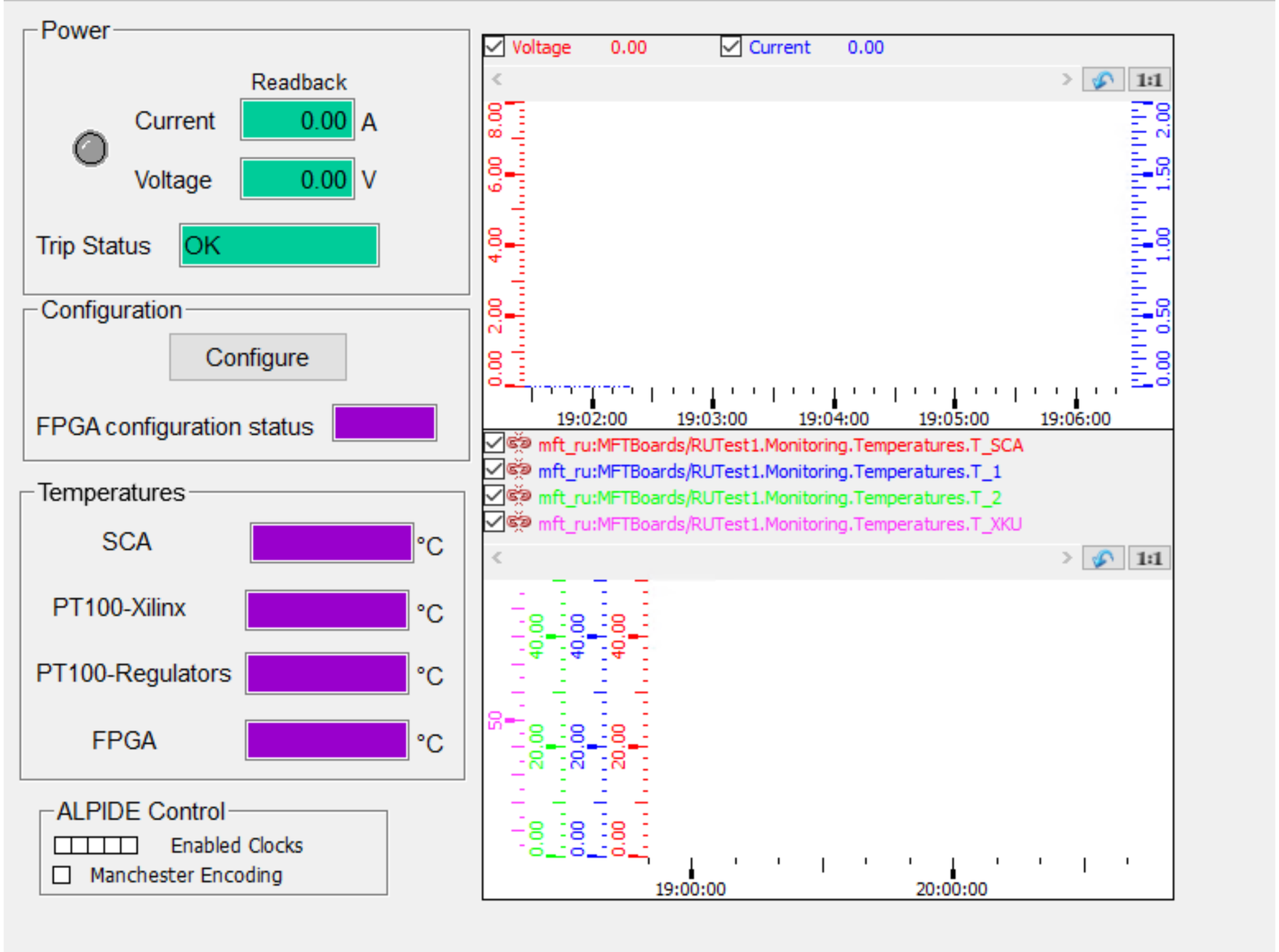}
  \caption{DCS panel of a single RU control for standard users.}
  \label{fig:RU_control_shifter}
\end{figure}

\begin{figure}[htb]
  \centering 
  \includegraphics[width=0.6\linewidth]{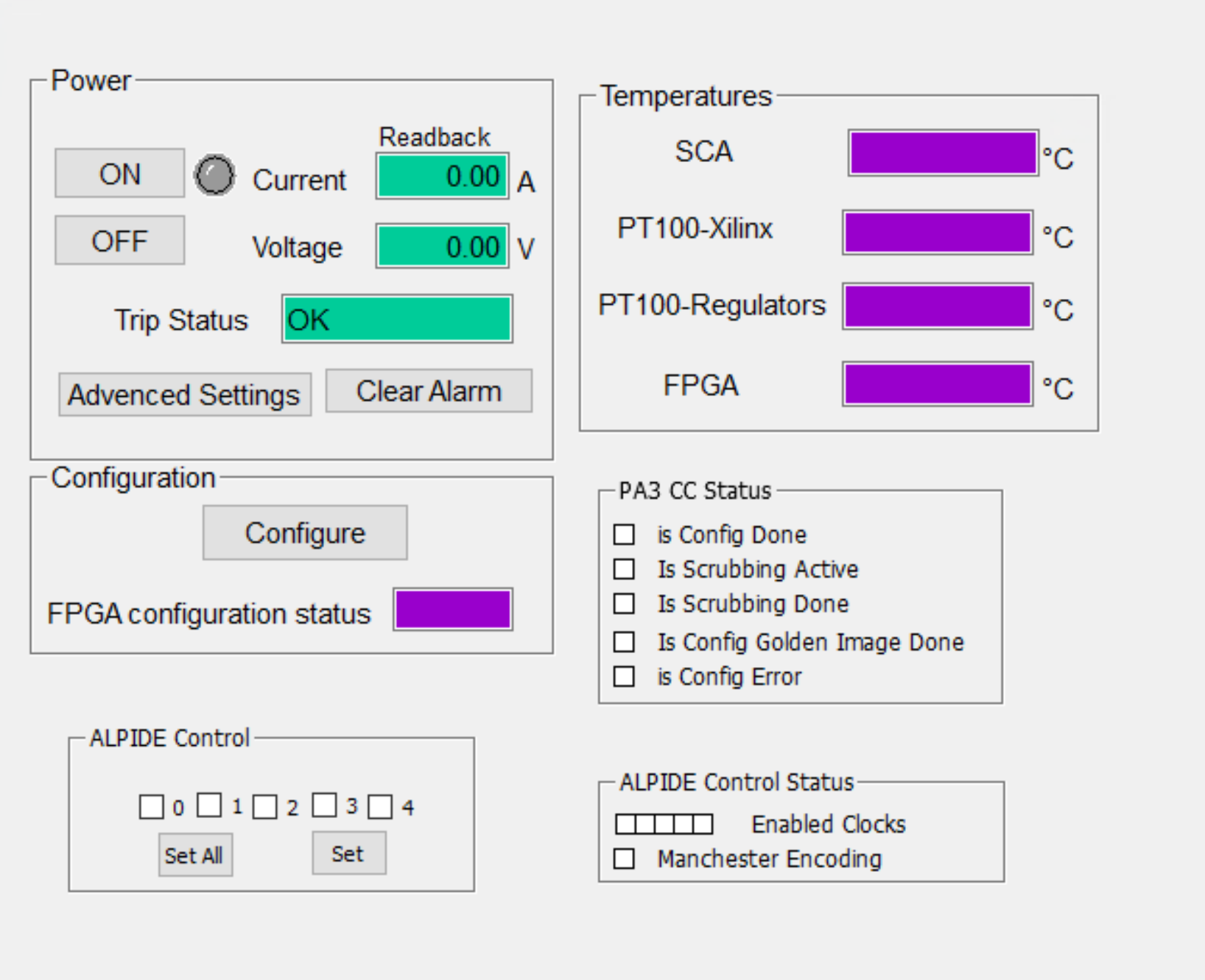}
  \caption{DCS panel of a single RU control for experts.}
  \label{fig:RU_control_expert}
\end{figure}

Separate DCS panels are developed to be used by standard users and by experts.
The DCS panels of a single RU control for standard users and experts are shown as examples
in Figs.~\ref{fig:RU_control_shifter} and \ref{fig:RU_control_expert}, respectively.
Standard users are basically restricted to monitor the RU conditions.
Only experts are allowed to turn on and off the power of the RU
and to set the voltage values using the advanced setting functions.
The FPGA configuration is easily accessible via a dedicated button.
The temperature values of the RU board and chips on it are used
to activate a software interlock if any temperature value exceeds the threshold.
The GUI is organized in a tree-like structure,
and the panels at a higher hierarchical stage
can control all RUs simultaneously.

%% file: Summary.tex
\section{Summary and Outlook}
\label{sec:summary}
The MFT is a new silicon pixel detector installed in the ALICE experiment
for Runs 3 and 4 of the LHC at CERN, starting in 2022,
in order to improve the muon tracking capability at forward rapidity.
The MFT DCS has been developed within the frameworks of CERN and the ALICE DCS\@. 
It controls and monitors the low voltage power supplies,
the detector, the readout modules, and the cooling system of the MFT\@.
The FSM is a key element both for operation of the detector
and for the software interlock.
In addition, the DSS serves as the ultimate hardwired interlock.
The DCS is implemented, used, and tested in the quality assurance of the MFT ladders.
It works perfectly at the smoke test of about 500 ladders.
It is also integrated in the surface commissioning of the MFT,
where its all functionalities are tested.
The MFT DCS system is operational,
constantly improving and adding needed functionalities,
and ready for the real detector operation.

Interlock scenarios to be implemented on the FSM and/or DSS
will be defined for specific alert cases.
The DCS with full functionalities will be implemented to the ALICE detector
and installed in the cavern for the physics runs.

%% file: Acknowledgments.tex
\acknowledgments
We appreciate the ALICE MFT collaboration
for their support in the design and implementation of the detector control system.
We are grateful to the DCS and O$^2$ teams of the ALICE collaboration for their technical help,
and to the entire ALICE collaboration.
This work was in part supported by JSPS KAKENHI grant numbers JP15H03664 and JP18H05401,
and by the Toshiko Yuasa France Japan Particle Physics Laboratory (TYL-FJPPL) project HAD\_02.